\documentclass[11pt]{article}
\usepackage{moriond,epsfig}
\usepackage{caption}
\usepackage{mathtools}
\usepackage{multirow}

\def\Journal#1#2#3#4{{#1} {\bf #2}, #3 (#4)}

\def\PRD{{\em Phys. Rev.} D}

\def\CQG{\em Class. Quantum Grav}

\def\NJP{\em New J. Phys.}
\def\mco{\multicolumn}

\def\be{\begin{equation}}
\def\ee{\end{equation}}
\def\bea{\begin{eqnarray}}
\def\eea{\end{eqnarray}}

\begin{document}
\vspace*{4cm}
\title{CONSTRAINING THE DISTANCE TO INSPIRALING NS-NS WITH EINSTEIN TELESCOPE}

\author{I. KOWALSKA-LESZCZYNSKA$^1$, 
          T. BULIK$^{1}$}

\address{$^1$Astronomical Observatory, University of Warsaw, Al Ujazdowskie 4, 00-478 Warsaw, Poland\\}

\maketitle
\abstracts{
}
\section{Introduction}
Einstein Telescope (ET) is a planned  third generation gravitational waves detector located in Europe\cite{Punturo2010}. Its design will be different from currently build interferometers:
First, ET will be located underground in order to  reduce the seismic noise. The arms length will be 10 km, and  the configuration of arms will be different from all interferometers build so far i.e. 
there will be three tunnels in a triangular
shape. ET will consist of three interferometers rotated by a 60$\deg$ with respect to each other in one plane. 
One of the biggest challenges for ET will be to determine sky position and
distance to observed sources. If an  object is observed in a few interferometers simultaneously one can estimate the position using traingulation from time delays\cite{Fairhurst2009},
but so far there are no plans for a network of third
generation detectors. Another possibility to deal with that problem is by using multimessenger approach, because redshift and sky position could be recovered from electromagnetic observations.
However,  in most cases of ET detection there will be only gravitational signal.
In this paper we  present  a novel method of estimating distance and position in the sky of merging binaries. While our  procedure is not as accurate as the multimessenger method, it
can  be applied  to all observations, not just the  ones with electromagnetic counterparts.

\section{Distance estimation using one interferometer}
For simplicity let us consider the case of observation of a double neutron star.
In gravitational waves we will be observing directly two quantities: signal to noise ratio ($\rho$), which is a complicated function of the source properties, as well as 
the detector characterization, and redshifted chirp mass ($M_z=(1+z)M_{chirp}$, $M_{chirp}=(M_1 M_2)^{3/5} (M_1+M_2)^{-1/5}$). In this particular case we consider only binaries consisting of two 
neutron stars of equal masses $M_1=M_2=1.4$ M$_\odot$, so $M_{chirp}=1.2$ M$_\odot$.
The signal to noise ratio in the  quadruple approximation for merging double compact objects is well known \cite{Taylor2012a}:
\begin{equation}
\label{eq:snr}
\rho \sim \frac{\Theta}{d_L(z)} (M_{z})^{5/6} \sqrt{\xi(z)},
\end{equation}
where $d_L$ is the luminosity distance, $M_{z}$ is the redshifted chirp mass, $z$ is the redshift, $\Theta$ is a  function of sky position and orientation of the source,
and $\xi$ is the function that determines fraction of the sensitivity window filled by a signal (it  depends on the chirp mass, and for NSNS binaries its value is close to unity).
For a given binary that will be observed in the detector, we can measure $\xi$ directly, by measuring the cutoff frequency when the inspiral ceases.

The function $\Theta$ depends on the sky position of the source  $\Omega(\vartheta, \varphi)$ and on the orientation of the orbit with respect to
the line of sight $\Omega_p(\Psi, i)$:
\begin{equation}
\begin{multlined}
\Theta=2 \sqrt{(1+\cos^2{i}) (F_{+})^2 + 4 \cos^2{i} (F_{x})^2},\\
F_{+}=0.5(1+\cos^2{\vartheta}) \cos{2\varphi} \cos{2\Psi}-\cos{\vartheta} \sin{2\varphi} \sin{2\Psi}, \\
F_{x}=0.5(1+\cos^2{\vartheta}) \cos{2\varphi} \sin{2\Psi}+\cos{\vartheta} \sin{2\varphi} \cos{2\Psi}. \\
\end{multlined}
\end{equation}
%\clearpag
The density  of sources in a unit volume  can be expressed by: 
\begin{equation}
\frac{d^2n}{dzd\Omega d\Omega_p}=\frac{n(z)}{1+z} \frac{dV}{dz}.
\label{eq:dndzdO}
\end{equation}
The comoving volume is $\frac{dV}{dz}=4 \pi \frac{c}{H_0} \frac{r^2(z)}{E(\Omega,z)}$, and $E(\Omega,z)=\sqrt{\Omega_\Lambda+\Omega_M(1+z)^3}$.
Then we obtain for a single detector
\begin{equation}
\begin{multlined}
\frac{dn}{dz} = \int{d\Omega d\Omega_p \frac{n(z)}{1+z} \delta(\rho-\rho_m)}\\
=4\pi \frac{n(z)}{1+z} \frac{c}{H_0} \frac{r^2(z)}{E(\Omega,z)}\frac{d_L}{8r_0} \left(\frac{1.2}{M_z}\right)^{5/6}\frac{1}{\sqrt{\xi}}\\
\times P \left(\frac{\rho^m}{8r_0 (\frac{Mz}{1.2})^{5/6} \sqrt{\xi}}d_L(z)\right),\\
\end{multlined}
\end{equation}
where $n(z)$ is the merger rate, $r_0$ is the characteristic distance for a given detector (see Table 1 in paper by Taylor\cite{Taylor2012b} for more details), $\rho^m$ is the actual signal-to-noise ratio measured in the detector.

\section{Distance estimation using three co-located interferometers}
Design of the Einstein Telescope assumes three co-located interferometers lying in the same plane, so the  methods for distance estimation based on triangulation will not be possible. However,
a single source will be observed by each of the interferometer with a different orientation. There will be three different measurements of signal to noise ratio. That will
provide additional information about the observed source and it allows to constrain the 
distributions obtained in previous section.

The density of sources per unit volume given by Eq. \ref{eq:dndzdO} has to be integrated taking into account that we have three conditions to satisfy.We assume that each signal to noise ratio is measured with perfect accuracy:
\begin{equation}
\begin{multlined}
\frac{dn}{dz}=\int{d\Omega d\Omega_p \frac{n(z)}{1+z} \delta(\rho_1-\rho_1^m) \delta(\rho_2-\rho_2^m) \delta(\rho_3-\rho_3^m)}\\
=\int\frac{d^2n}{dzd\Omega d\Omega_p}4\pi \frac{n(z)}{1+z} \frac{c}{H_0} \frac{r^2(z)}{E(\Omega,z)}\frac{d_L}{8r_0} \left(\frac{1.2}{M_z}\right)^{5/6}\frac{1}{\sqrt{\xi} (\rho_1^m)^2}\\
\times \delta \left(\Theta_1 - \frac{\rho_1^m}{8r_0 (\frac{Mz}{1.2})^{5/6} \sqrt{\xi}}d_L(z)\right) \delta \left(\frac{\Theta_2}{\Theta1}-\frac{\rho_2^m}{\rho_1^m}\right) \delta \left(\frac{\Theta_3}{\Theta1}-\frac{\rho_3^m}{\rho_1^m}\right),\\
\end{multlined}
\end{equation}

For illustration we present four cases of binary neutron stars simulated  ET observations. The physical parameters of those sources, as well as the observed quantities are shown
in Table \ref{tab:nsns}. 

\begin{table}[t]
\caption{Physical parameters and observed quantities of four sources. For first three of them (A, B, C) sky position and orientation were chosen randomly from uniform distributions on ththe obtained signal to noise ratio (the binary is optimally oriented).}
\label{tab:nsns}
\vspace{0.4cm}
\begin{center}
\begin{tabular}{|c|c|c|c|c|c|c|c|c|c|c|}
\hline
&\mco{6}{|c|}{Physical parameters} &\mco{4}{|c|}{Observed quantities} \\
\hline 
&M$_1$= M$_2$ [M$_\odot$] & z & $\vartheta$ [rad] & $\phi$ [rad] & $\Psi$ [rad] & i [rad]& M$_z$ [M$_\odot$] & $\rho_1^m$ & $\rho_2^m$ & $\rho_3^m$ \\
\hline
A &\multirow{4}{*}{1.4} & 0.1 & 0.53 $\pi$ & 0.82 $\pi$ & 1.30 $\pi$ & 0.70 $\pi$ & 1.34 & 44.04 & 94.42 & 55.95 \\
B & &  0.5 & 0.71 $\pi$ & 0.18 $\pi$ & 1.38 $\pi$ & 0.08 $\pi$ & 1.83 & 41.46 & 42.42  & 45.35\\
C & &  1.0 & 0.34 $\pi$ & 1.61 $\pi$ & 0.64 $\pi$ & 0.66 $\pi$ & 2.44 & 9.99 & 10.57 & 12.69 \\
D & &  1.0 & 0 & $\pi$ & $\pi$ & 0 & 2.44 & 36.90 & 36.90 & 36.90\\
\hline  
\end{tabular}
\end{center}
\end{table}

\begin{figure}
\centering
\includegraphics[scale=0.9]{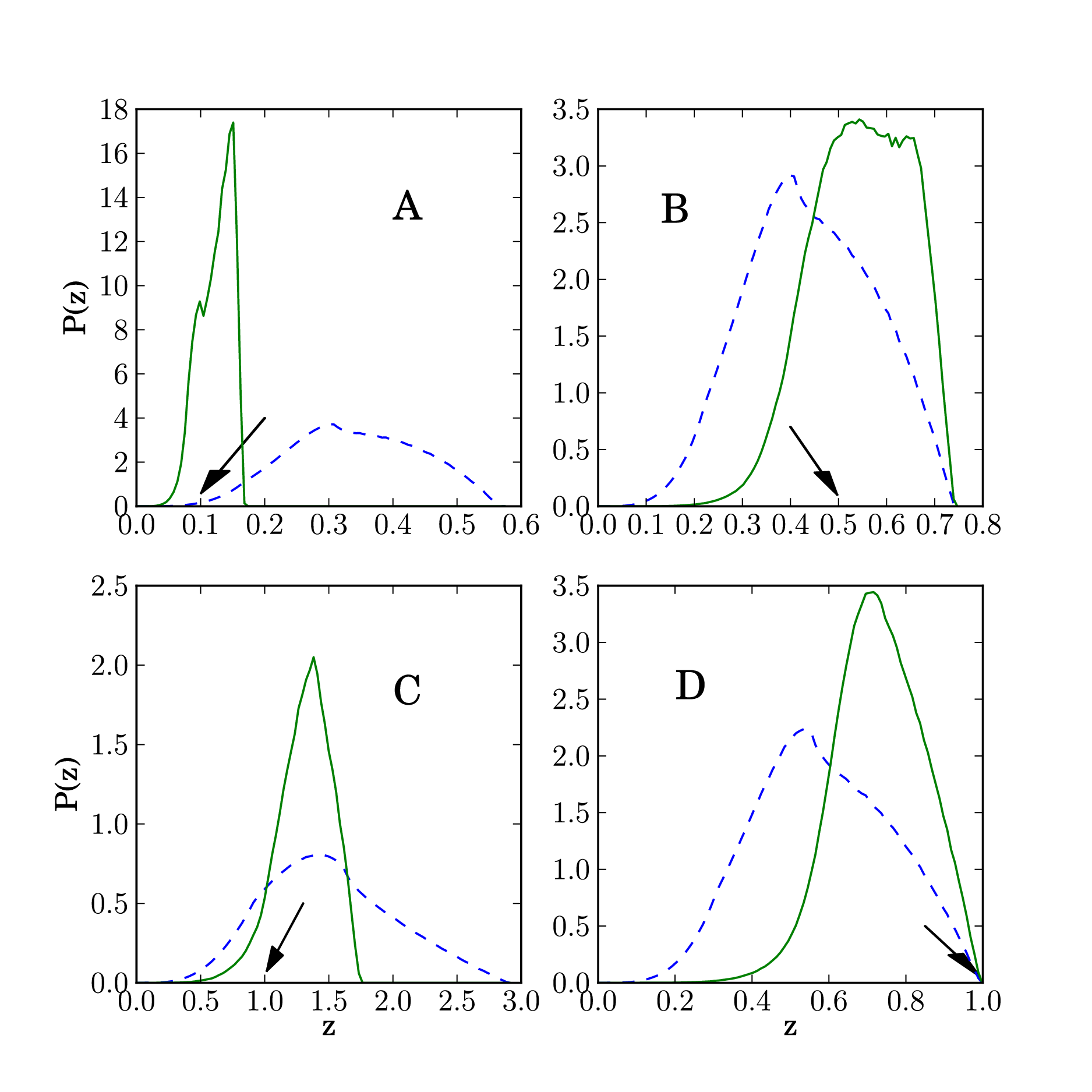}
\caption{Normalized redshift distributions for four NSNS system listed in Table \ref{tab:nsns}. Dashed line represent distribution obtained using only one interferometer,
while solid line represent distribution taking into account information from all three interferometers.The arrows indicate the position of the source.}
\label{rys:zDist}
\end{figure}

The results are shown in Fig. \ref{rys:zDist}. It can be clearly seen that 
our method can constrain distances to with the accuracy of about 20\%.

\section{Summary}
Distance measurements to merging binaries will be very challenging in the third generation detectors era. 
So far, there are no plans for any other detector than Einstein Telescope. 
In this paper, we presented a method that can be used to constrain distance distribution 
for a given double neutron star observation.
We have shown that it is possible to significantly improve  distance estimates 
using the measurements of the signal to noise ratio from all three interferometers .

\section*{Acknowledgments}
 This work was supported by the following Polish NCN  grants DEC-2011/01/V/ST9/03171 %\newline
and 2014/15/Z/ST9/00038.

\end{document}